\documentclass[12pt]{article}
\usepackage{amsfonts}
\usepackage{amsmath}
\usepackage[all]{xypic}

\csname @addtoreset\endcsname{equation}{section}
\newcommand{\eq}{\begin{equation}}
\newcommand{\feq}{\end{equation}}
\newcommand{\eqn}{\begin{eqnarray}}
\newcommand{\feqn}{\end{eqnarray}}
\newcommand{\arr}{\begin{eqnarray*}}
\newcommand{\farr}{\end{eqnarray*}}

\font\mybb=msbm10 at 12pt
\def\bb#1{\hbox{\mybb#1}}

\def\bR {\bb{R}}

\begin{document}

\begin{titlepage}
\begin{flushright}
Bicocca--FT--05--02\\
IFUM--821--FT
\end{flushright}
\vskip 15mm
\begin{center}
\renewcommand{\thefootnote}{\fnsymbol{footnote}}
{\Large \bf Plane waves from double extended spacetimes}
\vskip 18mm
{\large \bf {Sergio L.~Cacciatori$^{1,4}$\footnote{cacciatori@mi.infn.it},
Giovanni Ortenzi$^{2}$\footnote{giovanni.ortenzi@unimib.it} and
Silvia Penati$^{3,4}$\footnote{silvia.penati@mib.infn.it}}}\\
\renewcommand{\thefootnote}{\arabic{footnote}}
\setcounter{footnote}{0}
\vskip 10mm
{\small
$^1$ Dipartimento di Matematica dell'Universit\`a di Milano,\\
Via Saldini 50, I-20133 Milano, Italy. \\

\vspace*{0.5cm}

$^2$ Dipartimento di Matematica ed Applicazioni, Universit\`a
Milano Bicocca,\\
via  R. Cozzi, 53- I-20126 Milano, Italy. \\

\vspace*{0.5cm}

$^3$ Dipartimento di Fisica, Universit\`a degli studi di Milano-Bicocca, \\
piazza della Scienza 3, I-20126 Milano, Italy.\\

\vspace*{0.5cm}

$^4$ INFN, Sezione di Milano,\\
Via Celoria 16,
I-20133 Milano, Italy.\\

}
\end{center}
\vspace{2cm}
\begin{center}
{\bf Abstract}
\end{center}
{\small 
We study exact string backgrounds (WZW models) generated by nonsemisimple
algebras which are obtained as double extensions of generic D--dimensional 
semisimple algebras. We prove that a suitable change of coordinates always 
exists which reduces these backgrounds to be the product of the nontrivial
background associated to the original algebra and two dimensional Minkowski. 
However, under suitable contraction,
the algebra reduces to a Nappi--Witten algebra and the corresponding spacetime 
geometry, no more factorized, can be interpreted as the Penrose limit of the original
background. For both configurations we construct D--brane solutions and prove that
{\em all} the branes survive the Penrose limit. Therefore, the limit procedure can
be used to extract informations about Nappi--Witten plane wave backgrounds in
arbitrary dimensions.}

\end{titlepage}

\section{Introduction}

\label{intro}
One of the main topics in string theory is the construction of exact 
backgrounds,
that is two dimensional $\sigma$-models which are conformally invariant
at the quantum level and at all orders in the $\alpha'$ expansion.\\
A distinguished class of exact backgrounds is given by the WZW models 
on group manifolds where the vanishing of the
$\beta$ functions at all orders is ensured by the affine Sugawara construction
\cite{Mohammedi:1993rg}. 
For a WZW model associated to a given algebra,
the Sugawara construction exists if and only if the algebra possesses 
an ad--invariant, symmetric and non-degenerate metric.
In the case of a semisimple algebra this is the Killing metric and
all the renormalization effects give simply a correction to the metric. 
Generalizations to nonsemisimple cases have been extensively studied. 
In \cite{Nappi:1993ie} a WZW model based on the central extension of the
2d Poincar\'e algebra was constructed (NW model). This construction was then extended to more
general nonsemisimple cases \cite{Olive:1993hk}--\cite{Sadri:2003ib}, 
while the corresponding generalized
Sugawara quantization was completely analyzed in \cite{FFS:1994}
for the whole class of nonsemisimple algebras which admit an invariant metric, i.e. 
algebras obtained as double extensions of abelian or 
semisimple algebras \cite{MER}.

A second class of string backgrounds can be obtained by means of the Penrose
limit (\cite{Blau:2002mw},\cite{Russo:2002qj}). The two classes partially 
overlap since generalized NW models
can be obtained as Penrose limits of suitable geometries \cite{Stanciu:2003bn}.

In the present paper we investigate the spacetime geometries which arise from WZW
models associated to the abelian double extension of a generic semisimple $D$--dimensional
Lie algebra. 
We first give a general proof that the extended algebra can be always reduced
to the direct sum of the original algebra and a bidimensional abelian algebra
\footnote{This fact was shown in a more abstract way in \cite{FFS:1994} for the general
case of nonreductive algebras.}.
The corresponding spacetime geometry is then in some sense trivial since
it reduces to the product of the original spacetime with two dimensional
Minkowski. 
However, what makes these constructions interesting is that by taking a 
suitable In\"on\"u-Wigner contraction \cite{IW} of the extended algebra,
the new algebra which emerges is a Nappi--Witten like algebra. Therefore,
the geometry described by the corresponding sigma model is no more trivial, being
a $(D+2)$--dimensional Nappi--Witten background. 
We show that it is the Penrose limit of the original model associated to the 
nonsemisimple algebra. 

An interesting question which emerges is whether in the process of contracting
the algebra (or equivalently of taking the Penrose limit on the corresponding sigma model) 
information is lost. 
To answer this question we study brane configurations in both cases and prove that all 
brane solutions we find in the contracted model correspond to the Penrose limit of
brane solutions of the double extended original model. Therefore, all the information
goes safely through the limit. 

The plan of the paper is the following. In Section 2 we recall some
basic facts about one dimensional double extensions and give the general proof of
the fact that double extensions of semisimple Lie algebras are somehow
trivial. Our proof can be easily generalized to
the case considered in \cite{FFS:1994}. Next we show how to perform a suitable
In\"on\"u-Wigner contraction of the double extended algebra to obtain a nontrivial 
generalized NW algebra.
In Section 3 we construct the corresponding WZW model and show that  
the contraction actually corresponds to a Penrose limit on the 
corresponding string background. We implement the affine Sugawara construction 
and compute the central charge of the model. 
In Section 4 we construct brane configurations for
both models, the one associated to the double extended algebra and the
one corresponding to the contracted algebra. In particular, we prove that the
contraction can be used to extract all possible informations about
the limit background. Many technical details are collected in two Appendices.


\section{The double extended algebra and its contraction}

We consider a D-dimensional Lie algebra ${\cal {A}}$ with generators $\tau_i \ , i=1,\ldots,D $
satisfying
\eqn
\left[ \tau_i , \tau_j \right] =f^k_{\ ij} \tau_k \ .
\label{A}
\feqn
Its one dimensional double extension is obtained by adding the new generators
$H$ and $H^*$ such that
\eqn
& & \left[ \tau_i , \tau_j \right] =f^k_{\ ij} \tau_k +f_{ij} H^*  
\nonumber \\
& & \left[ H , \tau_i \right] =f_i^{\ j} \tau_j
\label{algebra}
\feqn
while $H^*$ is an element of the center. Here $f_{ij}$ are antisymmetric matrices 
constrained by the Jacobi identities
\eqn
{f_{ij}}^k f_{kl} + {f_{jl}}^k f_{ki} +{f_{li}}^k f_{kj} =0 \ . \label{jacobi}
\label{jacobi1}
\feqn
Defining the new generators $\tau_\cdot :=H$, $\tau_* :=H^*$ the previous
algebra can be written as 
\eqn
\left[ \tau_I , \tau_J \right] =f^K_{\ IJ} \tau_K
\label{algebra2}
\feqn
where $I,J,K =1,\ldots ,D,\cdot,*$.
The indices are highered and lowered by the bi-invariant metric
\eqn
\Omega_{IJ} =\left(
\begin{array}{ccc}
a K_{ij} & 0 & 0 \\
0 & b & 1 \\
0 & 1 & 0
\end{array} \right)
\label{metric1}
\feqn
$K_{ij}$ being an invertible $ad$-invariant metric for $\cal {A}$. 
If $\cal {A}$ is semisimple one can take $K_{ij}= h_{ij}$ where $h_{ij}$
is the Killing metric of the semisimple algebra. We note that the constant $b$ in the metric
is totally arbitrary and, at the classical level, it could be consistently set to zero.\\
The condition (\ref{jacobi}) has a nice interpretation. On the external algebra $\Lambda^*$
over $\cal{A}$ one can define the external derivative 
\eqn
\delta :\Lambda^* \longrightarrow \Lambda^*
\feqn
as the operator which maps the $p$-form $\lambda$ into the $p+1$-form $\delta \lambda$ given by
\eqn
\delta \lambda (v_1 ,\ldots ,v_{p+1} )=\sum_{q<r=1}^{p+1} (-1)^{r+q}
\lambda(v_1 ,\ldots, {\hat v}_q ,\ldots ,{\hat v}_r ,\ldots ,v_{p+1}, [v_q ,v_r ] )
\feqn
where $v_i$ are vectors of $\cal{A}$ and the hat means exclusion. 
The coefficients $f_{ij}$
define a $\bR$-valued two form ${\cal F} :=\frac 12 f_{ij} \mu^i \wedge \mu^j $, where
$\left\{ \mu^i \right\}$ is the basis of $\cal {A}^*$ dual to $\left\{ \tau_i \right\}$. 
Therefore eq. (\ref{jacobi1}) can be rewritten as
\eqn
\delta {\cal F} = 0 \ .
\feqn
We concentrate on the case of ${\cal {A}}$ being a semisimple Lie algebra: 
From the Whithead's second lemma, the second cohomology class
is trivial, $H^2 ({\cal {A}}) =0$, and we can write
\eqn
{\cal F} =\delta \lambda
\feqn
for a given $1$-form $\lambda$. In components this condition reads 
\eqn
f_{ij} ={f_{ij}}^k \lambda_k \ .
\label{polarized}
\feqn
Therefore, in the case of ${\cal {A}}$ semisimple the constant 
matrices $f_{ij}$ entering the one--dimensional double extension are
constrained to have the form (\ref{polarized}).
We will say that the double extension of a semisimple Lie algebra is {\em polarized} by the
vector $\lambda_k$.

Since $f_{ij}$ are the coefficients of an exact two form there must exist a basis of 
the algebra which eliminates $f_{ij}$ in (\ref{algebra}). 
This basis can be easily found: If we introduce the combinations
\eqn
T_i := \tau_i +\lambda_i H^* \ ,\quad  Z:= H-\lambda^2 H^* -\lambda^i \tau_i 
\ , \quad Z^* := H^* \ ,
\label{nbase}
\feqn
where $\lambda^2 \equiv \lambda^i a K_{ij} \lambda^j$, 
the commutation rules reduce to
\eqn
\left[ T_i , T_j \right] =f^k_{\ ij} T_k
\label{atrivial}
\feqn
with $Z$ and $Z^*$ both in the center. 
Under this redefinition the invariant metric transforms as
$\Omega_b \longrightarrow \Omega_{\tilde b}$ with $\tilde b =b-\lambda^2$.
We will call (\ref{nbase}) the {\em trivializing} basis.
\\
We have proven that the one dimensional double extension $D({\cal {A}})$ of a semisimple Lie 
algebra always reduces to the direct sum of the original 
algebra and a bidimensional abelian algebra
\eqn
D({\cal {A}}) ={\cal {A}} \oplus \bR^2 \ .
\feqn
The two sectors (${\cal A}$ and $\bR^2$) are in fact 
orthogonal because of the particular structure of the invariant metric.
As a consequence, the manifold realized via
WZW construction will be the direct product of the semisimple group associated to ${\cal A}$ and the two dimensional flat minkowskian spacetime
\eqn
e^{D({\cal A})} =e^{\cal A} \otimes \bR^{1,1} \ .
\feqn

It is however interesting to consider the WZW construction corresponding to
an algebra obtained as In\"on\"u-Wigner contraction of (\ref{algebra}). 
To this purpose we start from an ansatz for
the metric slightly different from (\ref{metric1}) in order to end up with a 
well-defined, invariant metric after the contraction. We consider
the three--parameter family of invariant forms (for given $a$ and $b$ constant)
\eqn
\Omega(\xi, \sigma, \rho) =
\left(
\begin{array}{ccc} \xi aK_{ij} & 0 & 0 \\
0 & \sigma b & \rho \\
0 & \rho &  0
\end{array}
\right)
\feqn
and define the rescaled generators
\eqn
P_i := \alpha \tau_i \ , \quad T :=\alpha^2 H^* \label{IW} \ , \quad S := \xi H \ .
\label{rescale}
\feqn
They generate the one--parameter family of algebras ${\cal A}_{\alpha}$ given by  
\eqn
& & \left[ P_i , P_j \right] = \alpha f^k_{\ ij} P_k + f_{ij} T 
\nonumber \\
& & \left[ S , P_i \right] = f_i^{\ j} P_j
\label{palgebra}
\feqn
In particular the r.h.s. of the commutators do not depend on $\xi$.

Before the contraction, the products between the elements of the basis give
the nonvanishing results
\eqn
(P_i ,P_j ) =a\xi \alpha^2 K_{ij} \ , \quad (S,S)= \sigma \xi^2 b 
\ , \quad (S,T) = \rho \xi \alpha^2  \ .
\feqn
If we then choose
\eqn
\xi = \frac 1{\alpha^2} \quad ,  \quad \sigma = \alpha^4  \quad , \quad 
\rho = 1
\feqn
we find that with respect to the new basis the product is well defined for
$\alpha$ going to zero and, independently of the parameters, we have
\eqn
(P_i ,P_j ) =a K_{ij} \ , \quad (S,S)=b
\ , \quad (S,T) =1 \ .
\label{cmetric}
\feqn
Therefore, taking the contraction $\alpha \to 0$ we obtain the algebra 
\eqn
& & \left[ P_i , P_j \right] =f_{ij} T \nonumber \\
& & \left[ S , P_i \right] =\frac 1a K^{jk} f_{ik} P_j \equiv f_i^{\, j} P_j
\label{calgebra}
\feqn
with $T$ central and invariant metric (\ref{metric1}).
This is a Nappi-Witten algebra, therefore no more trivializable.

We note that the change of basis (\ref{nbase}) which trivializes the algebra $D({\cal {A}})$
becomes singular 
in this limit according to the fact that the Nappi-Witten algebra is nonseparable.
As we will see in Section 3, the relation between the trivial algebra (\ref{atrivial})
and the NW algebra (\ref{calgebra}) through the contraction corresponds to the fact that
the Penrose limit of Cartesian product spaces may generate nontrivial spacetimes.

Before closing this Section we mention the fact that a slightly different contraction can
be performed by starting with rescaled generators 
\eqn
P'_i := \alpha \tau_i \ , \quad T' :=\alpha^2 H^* \label{IW2} \ , \quad S' := \xi (H -
\frac{b}{2} H^*)\ .
\label{rescale2}
\feqn
In the limit $\alpha \to 0$ we still obtain the algebra (\ref{calgebra}) but with metric 
(\ref{metric1}) corresponding to $b=0$.

\section{The WZW model for the double extended algebra}
\label{sec:sigma}
We now construct the WZW model associated to the algebra (\ref{algebra}). 
We parametrize the group elements as
\eqn
g=g_{\cal A} e^{uH+vH^*} \ ,\qquad  \qquad  g_{\cal A} =e^{\theta^i \tau_i}
\label{element}
\feqn
being $g_{\cal A}$ an element of the group $e^{\cal A}$.
Using the general identity
\eqn
e^{-\theta} \partial_\alpha e^\theta =\int_0^1 e^{-x\theta} \partial_\alpha \theta e^{x\theta} dx
\feqn
we can compute the left current $J:=g^{-1} d g$ and find
\eqn
J=j^i \exp{(uF)}_i^{\ k} \tau_k +( d \theta^i j^*_i + dv) H^*
+ du H
\feqn
where $F$ is the matrix $F_i^{\ k}:= -{f_{\cdot i}}^k$. The current $j^i$ is the current 
of the unextended algebra $j^i := d \theta^k j_k^{\ i}$, with
\eqn
j_k^{\ i} := \sum_{n=0}^\infty \frac{1}{(n+1)!} \theta^{j_1}\ldots \theta^{j_n}
{f_{k j_1}}^{k_1} {f_{k_1 j_2}}^{k_2} \ldots {f_{k_{n-2} j_{n-1} }}^{k_{n-1}}
{f_{k_{n-1} j_n}}^{\ i} \ ,
\feqn
whereas $j^*$ is given by
\eqn
j^*_i :=\sum_{n=1}^\infty \frac 1{(n+1)!} \theta^{j_1}\ldots \theta^{j_n}
{f_{ij_1}}^{k_1} {f_{k_1 j_2}}^{k_2} \ldots {f_{k_{n-2} j_{n-1}}}^{k_{n-1}}
f_{k_{n-1} j_n} \ .
\feqn
This current takes a relatively simple expression in the abelian case 
\eqn
j_i^* =\frac 12 \theta^j f_{ij}
\feqn
and in the polarized case where it reduces to
\eqn
j_i^* =(j_i^{\ j} -\delta_i^{\ j}) \lambda_j \ .
\label{polarj} 
\feqn

The WZW action on group manifold is given by
\eqn
S ~=~ \frac{1}{4\pi} \int_{\partial \Sigma} d^2\sigma 
\Omega_{IJ} J_\alpha^I J^{\alpha J} ~+~ 
\frac{i}{12\pi} \int_{\Sigma} d^3\sigma \Omega_{KL} f_{IJ}^{\ K} \epsilon^{\alpha \beta
\gamma} J^{I}_{\alpha} J^{J}_{\beta} J^{L}_{\gamma}
\label{WZW1} 
\feqn
where $\Omega_{ij}$ is the metric (\ref{metric1}) for the double extended algebra.
Using the fact that $F\cdot K$ is antisymmetric as a consequence of the invariance
of $K$, we find
\eqn
\Omega_{IJ} J_\alpha^I J^{\alpha J} = aK_{ij} j_\alpha^i j^{\alpha j}
+b\partial_\alpha u \partial^\alpha u +2\partial_\alpha u
(\partial^\alpha M^i j_i^* +\partial^\alpha v)
\feqn
which gives a sigma model with metric
\eqn
G_{ij} &=& aG^{({\cal A})}_{ij} \nonumber \ ,\\
G_{uu} &=& b \nonumber \ ,\\
G_{uv} &=& 1 \nonumber \ ,\\
G_{ui} &=& j_i^* \ ,
\label{spacetime1}
\feqn
where $G^{({\cal A})}_{ij} =K_{lm} j_i^{\ l} j_j^{\ m}$ is the metric
associated to the unextended algebra.
In the same way, using the Jacobi identities, we find for the nonvanishing
components of the simplectic structure
\eqn
B_{ij} =B^{({\cal A})}_{ij} +\frac u2 f_{lm} j_i^{\ l} j_j^{\ m} \label{Bmunu}
\label{B1}
\feqn                        
where locally $d B^{({\cal A})} =H^{({\cal A})}$ with
\eqn
& & B^{({\cal A})} =\frac 12 B^{({\cal A})}_{ij} j^i \wedge j^j \ , \qquad
H^{({\cal A})} = \frac 16 H^{({\cal A})}_{ijk} j^i \wedge j^j \wedge j^k \ .
\feqn
The quantization of the model can be performed nonperturbatively by means of
the Sugawara construction  \cite{Sugawara:1967rw}, \cite{Sommerfield:1968pk},
\cite{FFS:1994}. Given the level--$k$ current algebra
\eqn
j_I (z) j_J (w) = -\frac {2k\Omega_{IJ}}{(z-w)^2} +f_{IJ}^{\quad L}
\frac {j_L (w)}{z-w} ~+~ {\rm reg.}
\feqn
the energy-momentum tensor takes the form
\eqn
T =L^{IJ}:j_I j_J : 
\feqn
where $L^{IJ}$ is the inverse of the matrix
\eqn
L_{IJ} =(-4k \Omega +h)_{IJ}
\label{Ttensor}
\feqn
and $h_{IJ}$ is the Killing form of the double extended algebra  
\eqn
h_{IJ} =-f_{IK}^{\quad L} f_{JL}^{\quad K} \ .
\feqn
In the polarized case, given the position (\ref{polarized}) it takes the form
\eqn
h_{IJ} =\left(
\begin{array}{ccc}
h_{ij} & \lambda^k h_{ki} & 0 \\
\lambda^l h_{lj} & \lambda^k h_{kl} \lambda^l & 0 \\
0 & 0 & 0
\end{array} \right)
\label{killing}
\feqn
where $h_{ij}$ is the Killing form of the algebra $\cal{A}$. We note that, even if we were
to start with a classical invariant metric (\ref{metric1}) with $b=0$, a nontrivial $b$ 
would get produced by the quantization procedure.

Using the Sugawara construction we can compute the central charge as
$c = -4k L^{IJ} \Omega_{IJ}$. In our case we find
\eqn
c=D+\sum_{n=1}^\infty Tr\left( \frac 1{4ak} K^{-1} h \right)^n
\feqn
where $D$ is the dimension of the double extended algebra, 
$D=2+dim\{ {\cal A} \}$. Given the particular structures of $\Omega_{IJ}$
and $h_{IJ}$ the central charge turns out to be independent of the
parameter $b$.
In particular for the abelian case we have $c=D$, 
whereas for ${\cal A}$ a semisimple Lie algebra ($K_{ij} = h_{ij}$)
\eqn
c=D +\frac {D-2}{4ak-1} = c({\cal A}) + 2 \ .
\label{centralcharge}
\feqn
We now search for the coordinate transformation corresponding to the 
change of basis (\ref{nbase}). In the new basis the generic element of the group
(\ref{element}) takes the form
\eqn
g= e^{\theta^i T_i} e^{u\lambda^j T_j} e^{uZ +(v-\lambda_k \theta^k)Z^*}
=e^{\Psi^i(\theta^j,u)T_i} e^{uZ +(v-\lambda_k \theta^k)Z^*}
\feqn
where, using the Baker-Campbell-Hausdorff formula
\eqn
\Psi^i(\theta^j,u) = \theta^i + u \lambda^i +\frac{u}{2} \theta^j \lambda^k f^i_{\ jk} 
+ \frac{u}{12} \theta^j \theta^k \lambda^l f_{kl}^{\ m} f_{jm}^{\ \ i} 
+ \frac{u^2}{12} \theta^j \lambda^k \lambda^l f_{jk}^{\ m} f_{ml}^{\ \ i} + \cdots \ .
\feqn

In terms of the new coordinates
\eqn
& & \Psi^i = \Psi^i (\theta^j,u) \cr
& & \tilde{u}= u \cr
& & \tilde{v}= v-\lambda_k \theta^k
\label{trivial}
\feqn
the sigma model can be easily found by following the previous calculations
where we set $\lambda =0$. In particular, the metric turns out to be
a diagonal block matrix (in this case $j^*_i =0$, see eqs. (\ref{polarj}, \ref{spacetime1})) 
and our solution completely factorizes as
\eqn
e^{\cal A} (\Psi^i) \otimes \bR^{1,1} (\tilde{u} ,\tilde{v}) \ .
\feqn
Therefore, the spacetime geometries described by WZW models associated
to double extended algebras are somehow trivial
extensions of the spacetimes associated to the original semisimple
algebra. 
However, as already mentioned, nontrivial backgrounds can arise by
means of suitable Penrose limits. This will be the subject of the next Section.

\section{The WZW model for the contracted algebra: The Penrose Limit}
\label{subs:penrose}

It is well--known \cite{Blau:2002mw} \cite{Stanciu:2003bn} \cite{Bianchi:2004vf} that the four ${\rm NW}_4$ 
and six dimensional ${\rm NW}_6$ plane wave
backgrounds arise as Penrose limits of ${\rm AdS}_2 \times {\rm S}_2$ and 
${\rm AdS}_3 \times {\rm S}_3$, respectively. In particular, for the second case, 
in \cite{Stanciu:2003bn}
it has been shown that at the level of the algebras this limit
can be interpreted as a group contraction, as a consequence of the existence of a
null one--parameter subgroup corresponding to a null geodesic of the invariant metric. 

We generalize this result to the whole class of models associated to double extended
algebras. Starting from the original algebra (\ref{algebra}) endowed with the metric
(\ref{metric1}) we define the inner product of two generators as $\langle \tau_I , \tau_J
\rangle = \Omega _{IJ}$. Correspondingly, we see that there exist null one--dimensional
subgroups generated by $K \equiv (H - \frac{b}{2}H^*)$ and $H^*$. 
Therefore, in analogy with the ${\rm AdS}_3 \times {\rm S}_3$ case, 
we might expect the contracted group to correspond to a WZW model which describes 
the Penrose limit of the original spacetime. In Section 1 we have discussed  
the In\"on\"u--Wigner contraction of our original algebra. Starting from the rescaled 
generators (\ref{rescale2}) and taking the $\alpha \to 0$ limit amounts to perform the contraction
along the null vector $K$. On the other hand, the rescaled generators (\ref{rescale})
correspond to a contraction along a vector which is a linear combination of 
$K$ and $H^{\ast}$ \footnote{Since the two null vectors are not orthogonal the linear
combination $K + \frac{b}{2}H^*$ which defines the direction of our contraction is not 
strickly a null vector.}. 
The two cases differ by the value of the parameter $b$ appearing in
the metric (\ref{metric1}) which in the first case is zero, whereas in the second case
is arbitrary. 

In any case, the contraction of the original double extended algebra 
gives rise to
a NW--like algebra which is known to correspond to a ${\rm NW_{D+2}}$ background. 
Being this background a plane wave it can be reasonably expected to be the Penrose limit
of a nontrivial background.    
 
We now elaborate on that. To prove that the contracted algebra actually
corresponds to the WZW model in the Penrose limit, we need prove that the model
constructed directly from the algebra (\ref{calgebra}) coincides with the 
Penrose limit of the model (\ref{WZW1}, \ref{spacetime1}, \ref{B1}).  

The WZW model associated to the contracted algebra (\ref{calgebra}) is known \cite{kehagias}, 
since the  
algebra is the $(D+2)$ dimensional generalization of the NW algebra.  It has the general
structure (\ref{WZW1}) where the invariant metric is of the form (\ref{metric1}). 
By parametrizing the generic group element as
\eqn
g= e^{ {\Theta^i} P_i}  e^{u S + v  T}
\feqn  
the corresponding sigma model describes a spacetime with metric 
\eqn
G_{IJ} =\left(
\begin{array}{ccc}
G_{ij} & G_{iu} & 0 \\
G_{ui} & b & 1 \\
0 & 1 & 0
\end{array} \right) \ .
\label{spacetime2}
\feqn 

We now consider the Penrose limit of the sigma model constructed in Section 2 for the 
original nonsemisimple algebra. 
  
Given the group element (\ref{element}) we perform the rescaling (\ref{rescale}) of the
generators
\eqn
g= e^{\frac {\theta^i}{\alpha} P_i}  e^{ \alpha^2 u S +\frac v{\alpha^2} T}
\feqn
and define the new coordinates 
\eqn
\Theta^i = \frac {\theta^i}{\alpha} \quad , \quad  
U=\alpha^2 u \quad , \quad  V=\frac v{\alpha^2} \ .
\label{newcoord}
\feqn
Performing this change of coordinates in the original current $J =g^{-1} dg$ 
we find
\eqn
J =\frac 1{\alpha} j^i (\alpha \Theta) e^{\frac{1}{\alpha^2} UF_i^{\, k} } P_k +
\left( \frac 1{\alpha^2} j^* (\alpha \Theta) +dV \right) T +dU S
\feqn
where
\eqn
\frac 1\alpha j^i (\alpha \Theta) =d \Theta^i +\frac {\alpha}2 \Theta^k f^i_{\ jk}
d\Theta^j 
+o(\alpha)
\feqn
and
\eqn
\frac 1{\alpha^2} j^* (\alpha \Theta) =\frac 12 \lambda_j \Theta^k f^j_{\ ik} 
d\Theta^i +o(1) \ .
\feqn
These quantities have a well--defined limit $\alpha \to 0$. Therefore, taking this
limit and computing the background metric according to the general prescription 
$\Omega_{IJ} J_{\alpha}^I J^{\alpha J} = G_{IJ} \partial_\alpha X^I \partial^\alpha X^J$, 
$X^I = (\Theta^i, U,V)$, we find 
\eqn
G_{ij} &=& a K_{ij} \ , \\
G_{UU} &=& b \ , \\
G_{UV} &=& 1 \ , \\
G_{Ui} &=& \frac 12 \lambda_j \Theta^k f^j_{\ ik} \ .
\feqn
This metric is of the form (\ref{spacetime2}) so proving that 
the sigma model associated to the contracted algebra is the Penrose limit of the
original sigma model.
 
To find the two form $B$ let us recall that it comes out from the bulk term
$S[B]\sim \int H =\int dB $.
Now using the family of metrics with the chosen parameters, one has
\eqn
H &=& \frac 16 f_{IJK} J^I \wedge J^J \wedge J^K =
\frac 16 f_{ijk} j^i \wedge j^j \wedge j^k +\frac 12 f_{ij\cdot} j^i \wedge
j^j \wedge du \cr
&=& \frac 16 \alpha a K_{kl} {f_{ij}}^l \frac{j^i}\alpha \wedge \frac{j^j}\alpha
\wedge \frac{j^k}\alpha + \frac 12 \lambda_k {f_{ij}}^k
\frac{j^i}\alpha \wedge \frac{j^j}\alpha \wedge dU
\feqn
and taking the limit
\eqn
B_{ij} =\frac U2 \lambda_k f^k_{\ ij} \ .  
\feqn                        
The equivalence between the contraction of the algebra and the Penrose
limit of the corresponding background can be investigated also at the quantum level 
by studying how the Sugawara construction works under the limit $\alpha \to 0$.
To this end we consider the family of models parametrized by $\alpha$ and compute the
corresponding central charge (see eq. (\ref{centralcharge})) 
\eqn
c_\alpha = D + \frac {\alpha^2(D-2)}{4ak - \alpha^2} \ .
\feqn
In the limit $\alpha \to 0$ we find $c_\alpha \longrightarrow D$ which is the
correct central charge for the conformal sigma model associated to the contracted
algebra \cite{Nappi:1993ie, kehagias}. 
This proves that the limit is consistent also quantistically.

\section{Boundary states and D-brane configurations}
\label{sec:brane}

We can embed D--branes in a string background by determining
boundary states which preserve conformal invariance. 
In the case of WZW solutions this translates into boundary
conditions which have to be satisfied by the currents of the model.

Following \cite{Stanciu:1998ak} we impose the gluing conditions
\eqn
J_I (z) +{M^J}_I \bar J_J (\bar z ) =0
\label{gluing}
\feqn
where ${M^J}_I$ are determined by requiring conformal invariance and
current algebra to be preserved. 
If $M$ is defined on the generators (\ref{algebra2}) as $M(\tau_I )=\tau_J {M^J}_I$,
then the previous conditions read \cite{Stanciu:1998ak}
\eqn
& & M^T L M =L \label{confinv} \\
& & [M(\tau_I ),M(\tau_J )]=M([\tau_I ,\tau_J ]) \ , \qquad M^T \Omega M =\Omega 
\label{currsimm}
\feqn
where $L$ is given in (\ref{Ttensor}) and $\Omega$ is the invariant metric of the 
extended algebra.
The first condition follows from requiring conformal invariance,
while the second one comes from imposing the invariance of the current algebra.
We note that, given the particular form of the matrix $L$, the condition (\ref{confinv}) 
can be substituted by
\eqn
M^T h M =h \label{confinv1}
\feqn
where $h$ is the Killing form.

The solutions to these equations fix the boundary states. They can be found for the cases
of polarized algebras (\ref{algebra}, \ref{polarized}), trivial algebras (\ref{atrivial}) 
and contracted algebras (\ref{calgebra}).
Details on the procedure
for solving the equations in the three cases are given in the Appendices, while here we report
only the results and discuss their physical interpretation.

As shown in Appendix \ref{app:solution}, in the case of a semisimple algebra
polarized by the vector $\lambda$ the constraints have solution
\eqn
M^I_{\ J} =\left(
\begin{array}{ccc}
N^i_{\ j} -\sigma \lambda^i \lambda_j \quad  & \lambda^k N^i_{\ k}
-\nu \lambda^i \quad  & \sigma \lambda^i \\
\sigma \lambda_j & \nu & -\sigma \\
\lambda_k N^k_{\ j} -\mu \lambda_j & \gamma & \mu
\end{array}
\right)
\label{M}
\feqn
where $\nu, \sigma, \gamma, \mu$ are real constants satisfying the following equations
\eqn
& & \sigma (\sigma \lambda^2 +\sigma b -2\mu) =0 \label{conditions} \ , \\
& & 1+\nu \sigma \lambda^2 -\sigma \lambda_i N^i_{\ j} \lambda^j +\sigma b\nu
+\sigma \gamma -\mu \nu =0 \label{conditions1} \ , \\
& & \lambda^2 (1+\nu^2 )-2\nu \lambda_i N^i_{\ j} \lambda^j +b\nu^2 +2\gamma \nu
=b \ . \label{conditions2}
\feqn
The solutions to these equations will depend on a free parameter (for example $\nu$). 
The matrix $N^i_{\ j}$ has to be an isometry of the Killing metric
\eqn
N^l_{\ i} h_{lm} N^m_{\ j} =h_{ij}
\label{Ncondition}
\feqn
and can be realized as an element of the original semisimple Lie group in the
adjoint representation
\eqn
N^i_{\ j} =\{ e^{\theta^k \sigma_k} \}^i_{\ j} \ , \qquad
\{ \sigma_k \}^i_{\ j} =f^i_{\ kj}
\label{Nsolution}
\feqn
Therefore the parameters $\theta^i$, together with the free parameter $\nu$, parametrize
the moduli space of the solutions.

The equations (\ref{conditions}--\ref{conditions2}) select two main classes of solutions:
\begin{itemize}
\item The one given by $\sigma=0$ which we will call {\em class 0}.
In particular, in this class $\mu \nu =1$. 
\item The one given by $\sigma \neq 0$ which we call {\em class $\sigma$}.
\end{itemize}

From the previous solutions it is easy to extract informations about the boundary
states for the WZW model associated to the algebra ${\cal A}$ in the 
trivializing basis (\ref{nbase}). In fact, performing the change of basis 
(\ref{nbase}) amounts to consider the original algebra where we have set $\lambda =0$. 
Since the solutions (\ref{M}, \ref{conditions}--\ref{conditions2}) have a smooth limit for 
$\lambda \to 0$ 
the boundary states for the trivial case can be easily obtained 
\eqn
M^I_{\ J} =\left(
\begin{array}{ccc}
~~~ N^i_{\ j} \quad  & 0 & 0 \\
0 & \nu & -\sigma \\
0 & \gamma & \mu
\end{array}
\right)
\label{M2}
\feqn
where $\nu, \sigma, \gamma, \mu$ satisfy 
\eqn
& & \sigma (\sigma b -2\mu) =0  \ ,\\
& & 1+\sigma b\nu
+\sigma \gamma -\mu \nu =0  \ , \\
& & b\nu^2 +2\gamma \nu =b \ . \label{conditions3}
\feqn

\vskip 15pt

Finally, we consider boundary states for the model associated to the contracted algebra
(\ref{calgebra}). In this case the solutions to the constraints (\ref{confinv},\ref{currsimm}) 
read (see Appendix B for details)
\eqn
{M_{\phantom{}_{(0)}}}^I_{\ J} =\left(
\begin{array}{ccc}
N^i_{\ j} & r^i \quad  & 0\\
0  & \nu \quad &  0 \\
-\nu r_l N^l_{\ j}  & -\frac \nu2 r^2  \quad & \nu
\end{array}
\right)
\label{Mcontracted2}
\feqn
where $N^i_{\ j}$ still satisfies (\ref{Ncondition}) and the choice of the constants is restricted 
by the following equations
\eqn
&& \nu^2 =  1 \nonumber \\
&& {f_{ij}}^* =\nu N^l_{\ i} N^k_{\ j} {f_{lk}}^* \ .
\label{condcontracted2}
\feqn
Here, again, ${f_{ij}}^* = \lambda^k f_{ijk}$.
As explained in Appendix B, the last equation requires $\lambda^k$ to be an eigenvector 
of $N$ with eigenvalue $\nu$ or $-\nu$. This can happen only for particular choices of $N$.\\

\subsection{Brane solutions}

If the previous solutions allow for a geometrical interpretation, they define D-brane
configurations in the given background. In order to identify them, one has to translate 
the gluing conditions on the chiral currents into boundary conditions on the fields.
As discussed in \cite{Stanciu:1999id,Figueroa-O'Farrill:1999ie,D'Appollonio:2004pm}, 
the gluing conditions (\ref{gluing}) coincide with boundary conditions
for the WZW model on group manifold only for configurations near the identity.
Therefore, solving (\ref{gluing}) amounts to determine D--brane configurations 
in the group manifold passing through the identity 
\footnote{More general solutions, determining boundary states in a 
neighbourhood of a generic point $g$ would first require the translation of the conditions
(\ref{gluing}) from conditions involving algebra--valued quantities to conditions
for group--valued quantities (see \cite{Stanciu:1999id}).}.   

We will concentrate on finding such configurations.
To this end we parametrize the group elements as in (\ref{element}).
The chiral currents evaluated near the identity are then
\eqn
& &  J ( z ) =g^{-1}  \partial g =  \partial \theta^i \tau_i 
+  \partial u H + \partial v H^*
\nonumber \\
& & \bar J (\bar z) =-\bar \partial g g^{-1} = -\bar \partial \theta^i \tau_i 
-\bar \partial u H -\bar \partial v H^*  
\feqn
which shows \cite{Stanciu:1998ak} that Neumann boundary conditions correspond to $J(z) = 
-\bar{J}(\bar{z})$, i.e. to positive
eigenvalues of $M$, whereas Dirichlet conditions correspond to $J(z) =\bar{J}(\bar{z})$,
i.e. to negative eigenvalues.
In particular, if $(-1)^D det M$ is positive we find odd dimensional D-branes, 
whereas D-branes are even dimensional in the opposite case. 

The problem of determining D--brane configurations near the identity
is therefore translated into the
spectral problem for the matrix $M$ \footnote{We note that the matrix $M$
acts on the currents as a right multiplication so that we have to consider
the spectral equation for {\it left eigenvectors}.}  
\eqn
v_J {M^J}_I =\omega v_I \ .
\label{spectral}
\feqn
From the condition (\ref{confinv1}) it follows ${\rm det} M = \pm 1$ 
and, as a consequence, the eigenvalues satisfy $|\omega| =1$.

We solve eq. (\ref{spectral}) for the different cases, polarized (\ref{algebra}, \ref{polarized}),
trivial (\ref{atrivial}) and contracted (\ref{calgebra}).

In the polarized case, given the particular structure (\ref{M}) for $M$, we have
\eqn
&& {\rm det} \left( M^I_{\ J}-w \delta^I_{\ J} \right) 
\nonumber \\
&& = {\rm det} (N^i_{\ j} -w \delta^i_{\ j}) \left[ (w-\nu+\sigma \lambda^2 )(w-\mu)
+\gamma \sigma + \mu \sigma \lambda^2 -\sigma \lambda_i \lambda^j
N^i_{\ j} \right] = 0 \ .
\feqn
Therefore, for both the $0$ and $\sigma$ classes, $(D-2)$ eigenvalues are determined
by the eigenvalues of the isometry matrix $N$ of the invariant metric
$h_{ij}$. From the condition (\ref{Ncondition}) it follows that ${\rm det}N = \pm 1$
so that the eigenvalues $\xi$ of $N$ satisfy $|\xi|=1$. 
If we call $\Xi_{i,\xi}$ the left eigenvector of $N$ corresponding to the generic 
eigenvalue $\xi$, the eigenvector for the matrix $M$ is
\eqn
v_\xi =\left( \Xi_{\xi, i} \ , \lambda^j \Xi_{\xi, j} \ , 0 \right) \ .
\feqn
The remaining eigenvalues depend on the specific class. We then determine them separately for
the two classes.

\vskip 8pt
\underline{CLASS 0} \\
The extra eigenvalues are $w=\nu$ and $w=\mu =\frac 1\nu$. In particular
they have the same sign so determining two extra Neumann or two extra Dirichlet
conditions.
The corresponding eigenvectors are 
\eqn
& & v_\nu =\left( 0 \ , \cdots \ , 0  \ , 1 \ , 0 \right) \ , \\
& & v_{\frac 1\nu} =\left( \lambda_i \ , \frac{\lambda^2 - b}{2} \ , -1 \right)
\feqn
where we have used (\ref{conditions2}). 
The D-brane configurations corresponding to this class of solutions depend on
the sign of ${\rm det}N$. If the matrix $N$ has an even 
number $2p$ of negative
eigenvalues, the boundary conditions describe $(D-1-2p)$-brane
for $\nu >0$ and $(D-3-2p)$-brane for $\nu <0$.
Similarly, if $N$ has $2p+1$ negative eigenvalues the boundary configurations are
$(D-2-2p)$-brane for $\nu >0$ and $(D-4-2p)$-brane for $\nu <0$.

\vskip 10pt
\underline{CLASS $\sigma$} \\
In this case the equation for the extra eigenvalues is
\eqn
w^2 +w[-\mu -\nu +\sigma \lambda^2 ] +\mu\nu +\gamma \sigma -\sigma
\lambda^i \lambda^j N_{ij} =0
\feqn
which, using (\ref{conditions},\ref{conditions1})
reduces to
\eqn
w^2 +w[-\mu -\nu +\sigma \lambda^2 ]-1 =0 \ .
\feqn
This equation has two distinct real solutions of opposite signs which give a
new Neumann and a new Dirichlet condition.\\
If $w_o \ , \ o=1,2$ are the solutions, then the corresponding eigenvectors
are
\eqn
v_o =\left( \lambda_i \ , \frac {w_o +\sigma \lambda^2 -\mu}\sigma \ , -1 \right) \ .
\feqn
The two sets of D-branes which are described by these solutions correspond to
the cases of even and odd number of negative eigenvalues for $N$. We have 
$(D-2-2p)$-brane and $(D-3-2p)$-brane geometries for even and odd number of
negative eigenvalues, respectively. 

Setting $\lambda = 0$ in the previous expressions we find the D--brane solutions
for the model based on the trivializing basis (\ref{nbase}). In particular, we note that
the structure of the eigenvectors becomes 
\eqn
&& v_\xi = \left( \Xi_{\xi, i} \ , 0 , 0 \right)
\nonumber \ , \\
&& v_o = \left( 0, \cdots , 0 , a , b \right)
\label{trivialvect}
\feqn
for suitable constants $a,b$, and consequently, $\langle v_\xi , v_o \rangle =0$. 

\vskip15pt

We now consider D--brane solutions for the model associated to the contracted algebra.
In this case the matrix of boundary conditions is given in (\ref{Mcontracted2}) 
and its spectral equation becomes
\eqn
{\rm det} \left( {M_{\phantom{}_{(0)}}}^I_{\ J}-w \delta^I_{\ J} \right) =
{\rm det} (N^i_{\ j} -w \delta^i_{\ j}) (w-\nu)^2 = 0 \ .
\label{spectraleq}
\feqn
We remind that in this case $\nu$ or $-\nu$ are required to be in the spectrum of $N$.

To find the eigenvectors we need discuss separately three cases: 
i) $\nu$ in the spectrum of $N$ with $\vec{\lambda}$ the corresponding 
eigenvector;
ii) $\nu$ in the spectrum of $N$ but $\lambda_j N^j_{\ i} = -\nu \lambda_i$;
iii) $\nu$ not in the spectrum of $N$. 

i) As first case we suppose $\nu$ to be eigenvalue of $N$ with degeneracy one
and $ \lambda_j N^j_{\ i}= \nu \lambda_i$. Then we have $(D-3)$ eigenvalues
$\xi \neq \nu$, $|\xi|=1$ of the matrix
$N$ plus the extra eigenvalue $\omega = \nu$ which will appear with degeneracy
three (see eq. (\ref{spectraleq})). The first $(D-2)$ eigenvectors are easy to find 
\eqn
&& v_\xi =\left( \Xi_{\xi, i} \ , \frac{\Xi_{\xi,j}r^j}{(\xi - \nu)} \ , 0 \right) 
\qquad \quad \xi \neq \nu \ ,
\nonumber \\
&& v^{(1)}_\nu =\left( 0 \ , \cdots \ , 0  \ , 1 \ , 0 \right) \ .
\label{evectors}
\feqn
We look for the two missing eigenvectors corresponding to the eigenvalue $\nu$.  
They will have necessarily the form $\left( Z_i\  , 0 \ , c \right) $ where 
the unknowns $Z_i$ and $c$ satisfy
\eqn
Z_j {N^j}_k -c\nu r_j {N^j}_k  =\nu Z_k \ , \label{zetazero2}\\
Z_i r^i -\nu \frac c2 r^2 =0   \label{zetazero}
\feqn
as follows from the requirement to be eigenvectors of $M$ with eigenvalue
$\nu$. The first equation is equivalent to
\eqn
c r^i =-N^i_{\ j} Z^j +\nu Z^i \label{cruciale} \ .
\feqn
This equation does not have solutions in general, unless a further constraint
is satisfied. Precisely, since we have supposed $ \lambda_j N^j_{\ i} = \nu \lambda_i$, 
it can be solved {\em iff} $\vec{r}$ is orthogonal to $\vec{\lambda}$. 
If this condition is satisfied, the system of equations (\ref{zetazero})
has solutions $Z_k =\lambda_k$ and $c=0$
and the corresponding eigenvector reads $v^{(2)}_\nu = (\lambda_i, 0,0)$.
The other eigenvector is obtained by solving (\ref{cruciale}) on the space
orthogonal to $\vec{\lambda}$. Here the solution exists uniquely. 
We can set for example 
$c=1$ and $Z_i$ satisfying (\ref{cruciale}). Note that, as a consequence, 
(\ref{zetazero}) is automatically satisfied.\\
ii) When $\vec{\lambda}$ satisfies $\lambda_j N^j_{\ i} = -\nu \lambda_i$ 
but $\nu$ is still in the spectrum of $N$ the two missing eigenvectors
exist {\em iff} $\vec{r}$ is orthogonal to $\vec{\Xi_\nu}$, where $ \vec{\Xi}_\nu N = \nu 
\vec{\Xi}_\nu$. They are given by 
$v^{(2)}_\nu=(\Xi_{\nu,i} , 0,0)$ and $v^{(3)}_\nu=(Z_i, 0,1)$ 
with $\vec{Z}$ solving (\ref{cruciale}).

iii) Finally, we suppose $\nu$ not in the spectrum of $N$. Then 
$v_\xi$ and $v^{(1)}_\nu$ in (\ref{evectors}) are $(D-1)$ eigenvectors. 
The last 
one is $(Z_i \ ,0 \ , 1)$, where $\vec Z$ is a solution of (\ref{cruciale}). 
Note that in this case $N^i_{\ j} -\nu \delta^i_{\ j}$ is invertible. 

To summarize, we have found that the extra eigenvectors determined by
eqs. (\ref{zetazero2}, \ref{zetazero}) always exist as far as $\nu$ is not in
the spectrum of the matrix $N$, whereas in the opposite case they exist if
and only if the vector $\vec{r}$ is orthogonal to the $\nu$--eigenvector.

\subsection{D--branes in the Penrose Limit}
\label{susec:sigma}

The eigenvectors found above can be used to embed D--branes into the
spacetime under consideration.
For simplicity we call Neumann vectors the eigenvectors corresponding
to positive eigenvalues. Given a boundary configuration $M$, we suppose
that $(p+1)$ Neumann vectors $v_\alpha \ , \ \alpha =0,\ldots,p$ 
are present. Therefore, they select the tangent directions to a $Dp$--brane. 

We consider the algebra generators  
\eqn
\sigma_\alpha =\sum_I v_{\alpha,I} \tau_I \quad , \quad \alpha =0,\ldots,p \ .
\feqn
Therefore, local coordinates $\chi^\alpha \ , \ \alpha =0,\ldots,p$
for the brane are related to the spacetime coordinates $X^I =(\theta^i , u, v)$ 
through the equation
\eqn
g= e^{\theta^i \tau_i} e^{uH+vH^*} = e^{\chi^\alpha \sigma_\alpha}  \label{embedding}
\feqn
which defines the embedding of the brane into the spacetime.\\
Using the Backer-Campbell-Hausdorff formula we find
\eqn
g=e^{\phi^i(\theta^j , u) \tau_i +uH +(v + \psi (\theta^j , u)) H^*}
\feqn
with
\eqn
&& \phi^i (\theta^j , u) =\theta^i -\frac u2 \theta^j f_j^{\ i} +\frac {u^2}{12}
\theta^j f_j^{\ k} f_k^{\ i} +\frac u{12} \theta^l \theta^j f_l^{\ k}   
f_{kj}^{\ \ i} +\ldots \cr
&& \psi (\theta^j ,u) = - \frac u{12} \theta^i \theta^j f_i^{\ k}f_{kj} +\ldots
\feqn
so that the embedding (\ref{embedding}) becomes
\eqn
&& \chi^\alpha v_{\alpha}^i = \phi^i (\theta^j , u) \ ,\cr
&& \chi^\alpha v_{\alpha}^{\cdot} =u \ , \cr
&& \chi^\alpha v_{\alpha}^{*} = v + \psi (\theta^j , u) \ .
\label{embedding2}
\feqn

This construction can be carried on for the models associated to the
double extended algebra both in the original basis (\ref{algebra}) and
in the trivializing one (\ref{nbase}), and for the contracted algebra 
(\ref{calgebra}). In particular, it is worth noting that in the 
trivializing case, given the particular structure (\ref{trivialvect})
for the eigenvectors, D--brane solutions fall into two orthogonal classes:
D--branes embedded in the D dimensional spacetime associated to the 
unextended algebra and D--branes in $\bR^{1,1}$.

An interesting topic we are going to investigate concerns the behavior of
the brane solutions under Penrose limit. In the previous subsection
we have given brane solutions both for the model associated to the 
nonsemisimple algebra and for the model associated to its contraction. 
Since we have shown that the contracted model corresponds to a Penrose
limit of the original one, the natural question which arises is whether the brane
configurations corresponding to the contracted algebra can be {\em all} obtained
as Penrose limit of the original configurations or part of them
are lost in this limit. 

Before entering the details of the discussion, we make a preliminary observation.
If we rewrite the group element $g$ in terms of the rescaled basis $(P_i , S, T)$ 
and coordinates (\ref{newcoord}), then eqs. (\ref{embedding2}) become
\eqn
&& \frac 1\alpha \chi^\alpha v_{\alpha}^i = \frac 1\alpha \phi^i (\alpha \Theta^j , U) \ ,\cr
&& {\alpha^2} \chi^\alpha v_{\alpha}^{\cdot} = U \ ,\cr
&& \frac{1}{\alpha^2} \chi^\alpha v_{\alpha}^{*} =V +\frac 1{\alpha^2}
\psi (\alpha \Theta^j , U)
\feqn
where the right hand side has a finite limit for $\alpha \rightarrow 0$.
Therefore, we can introduce new coordinates on the brane $Y^\alpha := \alpha^\zeta \chi^\alpha$
for some parameter $\zeta$ such that
\eqn
\lim_{\alpha \rightarrow 0} \left( \frac {v^i_\alpha}{\alpha^{\zeta+1}}
\ , \frac {v^{\cdot}_\alpha}{\alpha^{\zeta-2}} \ ,
\frac {v^*_\alpha}{\alpha^{\zeta+2}} \right) =(w^i_\alpha \ , w^\cdot_\alpha \ , w^*_\alpha )
\feqn
is finite and gives an eigenvector for the $M_{(0)}$ boundary matrix in the
Penrose limit configuration.
This result seems to indicate that boundary configurations for the contracted
model can be found as a limit of the configurations of the original model
even if {\em a priori} we do not expect $M_{(0)}$ to be in general obtained as a limit of some 
$M$ of the original model. 
However, this is exactly the case as we are now going to prove in details.

We consider the constraints (\ref{currsimm},\ref{confinv1}) for the
one--parameter family of algebras ${\cal A}_{\alpha}$ given in eq. 
(\ref{palgebra}).
We are interested in studying the solutions to the
constraints for finite $\alpha$ and compare the results obtained
when $\alpha \to 0$ with the solutions (\ref{Mcontracted2}--\ref{condcontracted2})
of the contracted case. 

If we still set $f_{ij} = f_{ij}^{\ k} \lambda_k$
the constraints from the first condition in (\ref{currsimm}) read
\eqn
& & 0=\lambda^k {f_{ki}}^j M^\cdot_{\ j}  \\ 
& & 0= \alpha {f_{ki}}^j M^\cdot_{\ j} +\lambda^j f_{kij}  M^\cdot_{\ *}  \\
& & M^i_{\ l} M^j_{\ m} \lambda^k f_{ijk} = \alpha {f_{lm}}^k M^*_{\ k} +
\lambda^k f_{lmk} M^*_{\ *}  \\
& & M^i_{\ \cdot} M^j_{\ m} \lambda^k f_{ijk} =
\lambda^k {f_{km}}^h M^*_{\ h}  \\
& & M^i_{\ *} M^j_{\ m} \lambda^k f_{ijk} = 0  \\
& & M^i_{\ *} M^j_{\ \cdot} \lambda^k f_{ijk} =0  \\
& & \alpha {f_{ij}}^k M^i_{\ l} M^j_{\ m} -\lambda^j {f_{ij}}^k
(M^\cdot_{\ l} M^i_{\ m} -M^\cdot_{\ m} M^i_{\ l}) = \alpha M^k_{\ j} {f_{lm}}^j
+M^k_{\ *} \lambda^j f_{lmj} \\
& & \alpha {f_{ij}}^k M^i_{\ l} M^j_{\ \cdot} -\lambda^j {f_{ij}}^k
(M^\cdot_{\ l} M^i_{\ \cdot} -M^\cdot_{\ \cdot} M^i_{\ l}) =-M^k_{\ j}
{f_{hl}}^j \lambda^h  \\
& & \alpha {f_{ij}}^k M^i_{\ l} M^j_{\ *} -\lambda^j {f_{ij}}^k
(M^\cdot_{\ l} M^i_{\ *} -M^\cdot_{\ *} M^i_{\ l}) =0  \\
& & \alpha {f_{ij}}^k M^i_{\ \cdot} M^j_{\ *} -\lambda^j {f_{ij}}^k
(M^\cdot_{\ \cdot} M^i_{\ *} -M^\cdot_{\ *} M^i_{\ \cdot}) =0 \ .
\feqn
Proceeding as in Appendix A we find 
\eqn
M(\alpha)^I_{\ J} =\left(
\begin{array}{ccc}
N^i_{\ j} -\sigma \lambda^i \lambda_j \quad  & 
\frac{1}{\alpha} (\lambda^k N^i_{\ k} - \nu \lambda^i )\quad  & \alpha \sigma \lambda^i \\
\alpha \sigma \lambda_j & \nu & -\alpha^2 \sigma \\
\frac{1}{\alpha} (\lambda_k N^k_{\ j} -\mu \lambda_j) & \gamma & \mu
\end{array}
\right)    \label{Malfa}
\feqn
where the matrix $N$ still satisfies (\ref{struttura}, \ref{Ncondition}) and can
be chosen as in (\ref{Nsolution}). The constants appearing in $M$ are constrained by the 
following equations
\eqn
& & \sigma (\sigma \lambda^2 +\alpha^2 \sigma b -2\mu) =0 \label{b1} \ ,\\
& & \frac{1}{\alpha} \left[ 1+\nu \sigma \lambda^2 -
\sigma \lambda_i N^i_{\ j} \lambda^j -\mu\nu \right] + \alpha \sigma b\nu
+\alpha \sigma \gamma =0 \label{b2} \ ,\\
& & \frac{1}{\alpha^2} \left[ \lambda^2( 1 + \nu^2 ) - 2 \nu \lambda^m N^k_{\, m} \lambda_k
\right] +b\nu^2 +2\gamma \nu =b \label{b3}
\feqn
which come from the extra conditions in (\ref{currsimm},\ref{confinv1}).

In general the matrix $M(\alpha)$ and the system of equations (\ref{b1}--\ref{b3}) do not have 
a well-defined limit for $\alpha \to 0$.
However we can expand the matrix elements in a power series of $\alpha$ as follows
\eqn
&& N^i_{\ j} ={N_{\phantom{}_{(0)}}}^i_{\ j}+ \alpha {N_{\phantom{}_{(1)}}}^i_{\ j}+\alpha^2 {N_{\phantom{}_{(2)}}}^i_{\ j} +\ldots \cr
&& \sigma = \sigma_{\phantom{}_{(0)}} +\alpha \sigma_{\phantom{}_{(1)}} +\alpha^2 \sigma_{\phantom{}_{(2)}} +\ldots \cr
&& \mu = \mu_{\phantom{}_{(0)}} +\alpha \mu_{\phantom{}_{(1)}} +\alpha^2 \mu_{\phantom{}_{(2)}} +\ldots \cr
&& \nu = \nu_{\phantom{}_{(0)}} +\alpha \nu_{\phantom{}_{(1)}} +\alpha^2 \nu_{\phantom{}_{(2)}} +\ldots \cr
&& \gamma = \gamma_{\phantom{}_{(0)}} +\alpha \gamma_{\phantom{}_{(1)}} +\alpha^2 \gamma_{\phantom{}_{(2)}} +\ldots \cr
&& \lambda^i = \lambda^i_{\phantom{}_{(0)}} +\alpha \lambda^i_{\phantom{}_{(1)}} +
\alpha^2 \lambda^i_{\phantom{}_{(2)}} +\ldots \ .
\label{exp}
\feqn
For $M^i_{\ .}$ to have a finite limit we find
\eqn
&& \lambda_{\phantom{}_{(0)}}^k {N_{\phantom{}_{(0)}}}^i_{\ k} -\nu_{\phantom{}_{(0)}} 
\lambda_{\phantom{}_{(0)}}^i =0  \label{n0}
\label{eigen1}
\feqn
where $\nu_{(0)}^2 = 1$ as a consequence of (\ref{Ncondition}) at lowest order.
It follows that $M^i_{\ .}$ takes the form
\eqn
M^i_{\ .} \equiv  r^i = 
{N_{\phantom{}_{(1)}}}^i_{\ k} \lambda_{\phantom{}_{(0)}}^k -\nu_{\phantom{}_{(1)}} \lambda_{\phantom{}_{(0)}}^i
+{N_{\phantom{}_{(0)}}^i}_k \lambda_{\phantom{}_{(1)}}^k  -\nu_{\phantom{}_{(0)}} \lambda_{\phantom{}_{(1)}}^i \ .\label{vi}
\feqn
Similarly, for $M^*_{\ j}$ to be well-defined in the limit we have
\eqn
&& \lambda_{\phantom{}_{(0)} k} {N_{\phantom{}_{(0)}}}^k_{\ j} -\mu_{\phantom{}_{(0)}} \lambda_{\phantom{}_{(0)j}} =0 
\label{eigen2}
\feqn
so that 
\eqn
M^*_{\ j} \equiv s_j = 
{N_{\phantom{}_{(1)}}}^k_{\ j} \lambda_{\phantom{}_{(0)} k} -\mu_{\phantom{}_{(1)}} \lambda_{\phantom{}_{(0)} j} +\lambda_{\phantom{}_{(1)} k}
{N_{\phantom{}_{(0)}}^k}_j -\mu_{\phantom{}_{(0)}} \lambda_{\phantom{}_{(1)} j} \ .
\feqn
Note that the compatibility of (\ref{eigen1}) with (\ref{eigen2}) requires 
\eqn
\mu_{(0)} = \nu_{(0)} \ .
\label{munu}
\feqn 
Now we concentrate on equations (\ref{b1}--\ref{b3}). Inserting the expansions (\ref{exp}) in (\ref{b1}) 
and taking the limit $\alpha \to 0$ we find 
\eqn
\sigma_{\phantom{}_{(0)}} (\sigma_{\phantom{}_{(0)}} \lambda_{\phantom{}_{(0)}}^2 -2\mu_{\phantom{}_{(0)}}) =0
\feqn
which implies $\sigma_{\phantom{}_{(0)}}=0$ or $\sigma_{\phantom{}_{(0)}} 
\lambda_{\phantom{}_{(0)}}^2 =2\mu_{\phantom{}_{(0)}}$.
Using (\ref{n0}, \ref{eigen2}, \ref{munu}) in (\ref{b2}) one finds 
$\mu_{\phantom{}_{(0)}} \nu_{\phantom{}_{(0)}} =1$ and the relation
\eqn
\mu_{\phantom{}_{(0)}} \nu_{\phantom{}_{(1)}} +\mu_{\phantom{}_{(1)}} \nu_{\phantom{}_{(0)}} 
-\sigma_{\phantom{}_{(0)}} \nu_{\phantom{}_{(1)}} \lambda_{\phantom{}_{(0)}}^2 +\sigma_{\phantom{}_{(0)}}
\lambda_{\phantom{}_{(0)} i} {N_{\phantom{}_{(1)}}^i}_j \lambda_{\phantom{}_{(0)}}^j =0 \label{nu1} 
\feqn
which with (\ref{struttura}) at first order in $\alpha$ can be used to show that
\eqn
s_j =-\nu_{\phantom{}_{(0)}} \left[ {N_{\phantom{}_{(0)}}}^l_{\ j} -\sigma_{\phantom{}_{(0)}} \lambda_{\phantom{}_{(0)} j} 
\lambda_{\phantom{}_{(0)}}^l \right] h_{kl} r^k \ .
\feqn
Finally we consider (\ref{b3}). It has a finite limit if $\lambda_{\phantom{}_{(0)} i} 
{N_{\phantom{}_{(1)}}^i}_j \lambda_{\phantom{}_{(0)}}^j =0$ and the nontrivial part of the equation becomes
\eqn
&& 2\lambda_{\phantom{}_{(1)}}^2 + \lambda_{\phantom{}_{(0)}}^2 \nu_{\phantom{}_{(1)}}^2 -2\nu_{\phantom{}_{(0)}}\lambda_{\phantom{}_{(0)}}^k 
{N_{\phantom{}_{(2)}}}^i_{\ k} \lambda_{\phantom{}_{(0)} i}
-2\nu_{\phantom{}_{(0)}}\lambda_{\phantom{}_{(1)}}^k {N_{\phantom{}_{(0)}}}^i_{\ k} \lambda_{\phantom{}_{(1)} i} \cr
&& -2\nu_{\phantom{}_{(0)}}\lambda_{\phantom{}_{(1)}}^k {N_{\phantom{}_{(1)}}}^i_{\ k} \lambda_{\phantom{}_{(0)} i}
-2\nu_{\phantom{}_{(0)}}\lambda_{\phantom{}_{(0)}}^k {N_{\phantom{}_{(1)}}}^i_{\ k} \lambda_{\phantom{}_{(1)} i}
+2\gamma_{\phantom{}_{(0)}} \nu_{\phantom{}_{(0)}} =0 \ . \label{last}
\feqn
From (\ref{struttura2}) to second order in $\alpha$ we find
\eqn
{N_{\phantom{}_{(1)}}}^k_{\ i} {N_{\phantom{}_{(1)}}}^l_{\ j} h_{kl} +({N_{\phantom{}_{(0)}}}^k_{\ i}{N_{\phantom{}_{(2)}}}^l_{\ j}
+{N_{\phantom{}_{(2)}}}^k_{\ i} {N_{\phantom{}_{(0)}}}^l_{\ j}) h_{kl} =0
\feqn
which used in $r^2 := h_{ij} r^i r^j$ and then inserted in (\ref{last}) gives
\eqn
r^2 +2\gamma_{\phantom{}_{(0)}} \nu_{\phantom{}_{(0)}} =0 \ .
\feqn
Collecting all the results the final form for the matrix $M$ in the limit $\alpha \to 0$ is 
\eqn
{M(\alpha \to 0)}^I_{\ J} =\left(
\begin{array}{ccc}
N^i_{\ j} -\sigma \lambda^i \lambda_j & r^i \quad  & 0\\
0  & \nu_{\phantom{}_{(0)}} \quad &  0 \\
s_j  & \gamma_{\phantom{}_{(0)}}   \quad & \nu_{\phantom{}_{(0)}}
\end{array}
\right) 
\feqn
where
\eqn
&& N^i_{\ j} \lambda^j =\nu_{\phantom{}_{(0)}} \lambda^i \\
&& 2\gamma_{\phantom{}_{(0)}} \nu_{\phantom{}_{(0)}} +r^2 =0 \\
&& \nu_{\phantom{}_{(0)}}^2=1 \\
&& r^k ( N^l_{\ j} h_{kl} -\sigma \lambda_k \lambda_j) +\nu_{\phantom{}_{(0)}} s_j =0  
\label{ristretto} \\
&& \sigma(\sigma \lambda^2 -2\nu_{\phantom{}_{(0)}})=0 \ .
\feqn

We have to compare this result with the solution 
(\ref{Mcontracted2}--\ref{condcontracted2})
for the contracted algebra. For $\sigma =0$ the two solutions 
coincide exactly. The case $\sigma \neq 0$ is also 
included since it can be traced back to the case $\sigma =0$ by observing that if
$N^i_{\ j}$ is an isometry then $N^i_{\ j} - \sigma \lambda^i \lambda_j$ is also 
an isometry whenever $\sigma \lambda^2 = 2 \nu_{\phantom{}_{(0)}}$. Note 
that $\vec \lambda$ is eigenvector of ${M_{\phantom{}_{(0)}}}^i_{\ j}$ with 
eigenvalue
$\nu_{\phantom{}_{(0)}}$ when $\sigma =0$, whereas it corresponds to eigenvalue 
$-\nu_{\phantom{}_{(0)}}$ when $\sigma \neq 0$.
 
Given the arbitrariness of the vector $r^i$, 
we can conclude that the class of solutions obtained in the Penrose limit
coincides with the class of solutions for the contracted algebra.
Therefore the Penrose limit seems to carry on all the informations and 
nontrivial background configurations can be generated from the trivial ones
by means of this limit.  \\
In order to give further support to this statement we study the behavior of the 
eigenvectors under the limit. 

We first consider the case $\sigma=0$.
For $\alpha$ finite, the eigenvectors of the matrix $M(\alpha)$ in (\ref{Malfa}) are 
\eqn
&& V_\xi = \left( \Xi_{\xi,i} \ , \frac{1}{\alpha} \lambda^i \Xi_{\xi,i} \ , 0 \right) \ ,
\label{D-2} \\
&& v_\nu =\left( 0 \ , \cdots \ , 0  \ , 1 \ , 0 \right)  \ ,\label{vecnu}\\
&& v_{\frac 1\nu} =\left( \frac{\lambda_i}{\alpha} \ , x \ , -1 \right)
\label{vecnuinv}
\feqn
where $\Xi_{\xi}$ are eigenvectors of $N$ with eigenvalues $\xi$ and $x$ satisfies
\eqn
(1-\nu^2) \left( 2 x - \frac{\lambda^2}{\alpha^2} +b \right) =0 \ .
\label{x}
\feqn
We make the assumption for one of the $\xi$ eigenvalues to have the form
\eqn
\bar{\xi} = \nu + O(\alpha^2) \ .
\label{xibar}
\feqn
This includes both the cases $\nu$ in or not in the
spectrum of $N$. 
 
We first concentrate on the eigenvectors $V_\xi$ with $\xi \neq \bar{\xi}$.
In order to study their behavior under the limit $\alpha \to 0$ we expand 
the quantities appearing in $M$ as in (\ref{exp}) and similarly
\eqn
&& \Xi_{\xi}^i =\Xi_{\phantom{}_{(0)}}^i +\alpha \Xi_{\phantom{}_{(1)}}^i
+\ldots \ , \\
&& \xi = \xi_{\phantom{}_{(0)}} +\alpha \xi_{\phantom{}_{(1)}} +\ldots \ .
\feqn
Note that 
${\lambda_{\phantom{}_{(0)}}}_i N^i_{\phantom{}_{(0) \, j}} = 
\nu_{\phantom{}_{(0)}} 
\lambda_{\phantom{}_{(0)} \,j}$, $\nu_{\phantom{}_{(0)}}^2 =1$.
From the condition for $\Xi_\xi$ to be an eigenvector of $N$, up to
the first order in $\alpha$  we obtain 
\eqn
&& {\Xi_{\phantom{}_{(0)}}}_i {N_{\phantom{}_{(0)}}^i}_j =
\xi_{\phantom{}_{(0)}} {\Xi_{\phantom{}_{(0)}}}_j \ ,\\
&& {\Xi_{\phantom{}_{(1)}}}_i {N_{\phantom{}_{(0)}}^i}_j +
{\Xi_{\phantom{}_{(0)}}}_i {N_{\phantom{}_{(1)}}^i}_j =
\xi_{\phantom{}_{(1)}} {\Xi_{\phantom{}_{(0)}}}_j +
\xi_{\phantom{}_{(0)}} {\Xi_{\phantom{}_{(1)}}}_j \ . \label{xiuno} 
\feqn
In particular, being $\xi_{\phantom{}_{(0)}} \neq \nu_{\phantom{}_{(0)}}$,
$ {\Xi_{\phantom{}_{(0)}}}_j$ is orthogonal to $ {\lambda_{\phantom{}_{(0)}}}_j$
($N$ is a unitary matrix).
Using (\ref{xiuno}) and (\ref{vi}) we find
\eqn
\frac 1\alpha \lambda^i \Xi_{\xi,i} = {\Xi_{\phantom{}_{(0)}}^j}
\left( {\lambda_{\phantom{}_{(1)}}}_j
+\frac {{\lambda_{\phantom{}_{(0)}}}_i 
{N_{\phantom{}_{(1)}}^i}_j}{\xi_{\phantom{}_{(0)}}-\nu_{\phantom{}_{(0)}}} 
\right) +o(1) 
= \frac {\Xi_{\phantom{}_{(0)}}^j r_j}{\xi_{\phantom{}_{(0)}}-\nu_{\phantom{}_{(0)}}} +o(1) \ .
\feqn
Therefore, in the limit we obtain the first set of eigenvectors in (\ref{evectors}).

We then look for the remaining eigenvectors. The eigenvector (\ref{vecnu}), being
independent of $\alpha$  survives the limit and coincides
with the extra eigenvector in (\ref{evectors}) corresponding to eigenvalue $\nu_{\phantom{}_{(0)}}$. 

Now the question is whether in the limit other $\nu_{\phantom{}_{(0)}}$ eigenvectors arise.
In order to answer, we first note that when $\xi = \bar{\xi}$ ($\bar{\xi}_{\phantom{}_{(0)}}
= \nu_{\phantom{}_{(0)}}$, $\bar{\xi}_{\phantom{}_{(1)}}= \nu_{\phantom{}_{(1)}}$) 
multiplying (\ref{xiuno}) by 
${\lambda_{\phantom{}_{(0)}}}_i$ we obtain the condition
\eqn
\nu_{\phantom{}_{(1)}} =0 \ .
\label{nu1cond}
\feqn
Inserted in (\ref{vi}) it says that $r^i$ has to be orthogonal
to ${\lambda_{\phantom{}_{(0)}}}_i$. This is exactly the condition we
found in the model associated to the contracted algebra for the existence 
of extra $\nu_{\phantom{}_{(0)}}$ eigenvectors. 

To obtain these eigenvectors in the limit we start considering
$V_{\bar{\xi}}$. Since
$\bar{\xi}_{\phantom{}_{(0)}} =\nu_{\phantom{}_{(0)}}$ we can take 
${\Xi_{\phantom{}_{(0)}}}_i ={\lambda_{\phantom{}_{(0)}}}_i$. As a consequence
the eigenvector $V_{\bar{\xi}}$ is generically divergent for $\alpha \to 0$. 
However, we can obtain a finite result by acting with the limit on the linear
combination 
\eqn
\eta \equiv V_{\bar{\xi}} -\frac 1{\alpha} \lambda_{\phantom{}_{(0)}}^2 v_\nu \ .
\label{eta}
\feqn 
This is a $\nu$ eigenvector for $M(\alpha)$ up to terms $O(\alpha^2)$ and in the limit
it gives rise to a $\nu_{\phantom{}_{(0)}}$ eigenvector for the contracted matrix. 
By a further finite subtraction, the resulting eigenvector can
be set into the form $v_\lambda =({\lambda_{\phantom{}_{(0)}}}_i \ , 0 \ , 0)$.

Finally we have the $1/\nu$ eigenvector (\ref{vecnuinv}) with $x$ solution of
(\ref{x}). From this equation expanded in powers of $\alpha$ 
we obtain again the condition (\ref{nu1cond})  and the extra constraint 
$\nu_{\phantom{}_{(2)}} =0$ \footnote{ This is not really an extra constraint 
but only a consequence of the choice $x$ instead of $x/\alpha$ in (\ref{vecnuinv})}.  
We consider the linear combination
$v_{\frac{1}{\nu}} - \frac{1}{\alpha} \eta$ with $\eta$ given in (\ref{eta}).
This is eigenvector of $M(\alpha)$ with eigenvalue $\nu_{\phantom{}_{(0)}}$, 
up to $O(\alpha)$ terms.
After a suitable finite subtraction, in the limit it generates 
$({\lambda_{\phantom{}_{(1)}}}_i \ , 0 \ , -1)$ which is the last $\nu_{\phantom{}_{(0)}}$ 
eigenvector of the contracted matrix. 
In this case from (\ref{vi}) it follows 
\eqn
r^i = {\lambda_{\phantom{}_{(1)}}^j} {N_{\phantom{}_{(0)}}^i}_j
-\nu_{\phantom{}_{(0)}} \lambda_{\phantom{}_{(1)}}^i \ .
\feqn
This is exactly the condition (\ref{cruciale}) with $c=-1$ found in the previous subsection
for the contracted case.\\
From this analysis we can conclude that in the case $\sigma=0$ 
we find a {\em complete} correspondence 
between the eigenvectors of $M(\alpha)$ for $\alpha \to 0$ and the eigenvectors
of $M_{\phantom{}_{(0)}}$.  \\
The case $\sigma \neq 0$ can be treated in a similar manner 
and there are no problems to prove that the limit exists in any case and gives
rise to the expected eigenvectors for the contracted matrix.
 
As remarked  at the beginning of this section, the algebraic method for finding
D--branes configurations holds only for D--branes passing through the identity of the group manifold $G$.
However the generic D--brane passing through a point $g \in G$ can be obtained, as shown in
\cite{Stanciu:1999id}, by a suitable ``pull-back'' to the origin of the gluing matrix which does not
affect the Penrose limit process. Accordingly, our result holds for a generic D--brane.
Therefore, we have proved that {\em all} the D--brane configurations of a 
(D+2)--dimensional NW background can be obtained as Penrose limit of D--brane solutions 
of the background associated to the double extension of a semisimple D--dimensional
algebra.

\section{Conclusions}
We have considered WZW models based on the double extension of a generic semisimple
algebra. We have given a general proof that the corresponding background is simply 
the cartesian product between the group manifold associated to the original semisimple 
algebra and a bidimensional Minkowski spacetime. However, less trivial spacetime configurations
can be obtained by taking a suitable Penrose limit. In this limit a generalized $(D+2)$--dimensional
Nappi--Witten background arises. We have shown that the Penrose limit corresponds at
the level of the algebras to a suitable In\"on\"u-Wigner contraction. In fact, the NW
background can be realized as a WZW model associated
to the double extension of an abelian algebra which is obtained as a
contraction of our original extended algebra.\\
We have considered brane states which can live in these spacetime backgrounds.
In particular, we have shown that non only the brane configurations of the double
extended model survive the Penrose limit, but {\em all} the algebraically defined brane states 
living in the generalized NW background can be obtained as such a limit.\\
We have argued that the correspondence between the Penrose limit
and the algebra contraction is consistent also at the quantum level. In fact, using the Sugawara
construction, we have computed the central charge of the one--parameter family of nonlinear sigma
models associated to the family of algebras ${\cal A}_{\alpha}$ and proved that in the limit $\alpha
\to 0$ they generate the correct central charge of the NW sigma model.  
Our results give evidence that the procedures of contraction and quantization schould
commute, as indicated by the following diagram\\
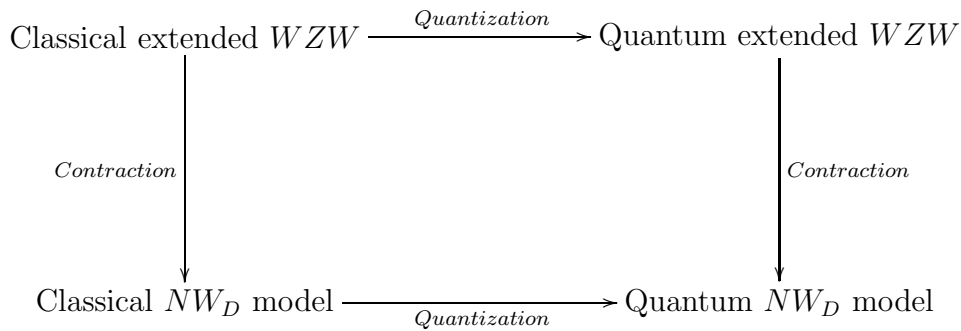
$$
\xymatrix{\mbox{Classical extended $WZW$} \ar[rrr]^{Quantization} \ar[ddd]_{Contraction} 
&&& \mbox{Quantum extended $WZW$} \ar[ddd]^{Contraction}\\ 
&&& \\
&&&\\
\mbox{Classical $NW_D$ model} \ar[rrr]_{Quantization} &&& \mbox{Quantum $NW_D$ model} }
$$
We expect this correspondence to be very general. It might be useful to gain 
informations on Nappi--Witten backgrounds starting from the original double
extended model. In particular, it could be used to construct the vertex operator algebras 
for the Nappi--Witten model $via$ the Kac formalism (\cite{WoInPr}) in order to reproduce 
the vertex operators found in \cite{D'Appollonio:2003dr}, \cite{D'Appollonio:2004pm} 
and \cite{Cheung:2003ym}.
\\
Possible generalizations of our results could be investigated for the cases of 
supersymmetric and/or noncommutative WZW models.

\section*{Acknowledgments}
\small

S.~C. and G.~O. would like to thank Bert Van Geemen for numerous helpful conversations.
We also thank Dietmar Klemm and Giuseppe Berrino for discussions
and Marco Rusconi for useful questions.\\
This work was partially supported by INFN, COFIN prot. 2003023852\_008 and the 
European Commission RTN program MRTN--CT--2004--005104 in which S.~C.~is
associated to the University of Milano--Bicocca.
\normalsize

\newpage

\begin{appendix}
\section{Solution to the constraints in the polarized case}
\label{app:solution}

We solve the constraints (\ref{currsimm},\ref{confinv1}) for $\cal{A}$ semisimple,
with the position (\ref{polarized}).
The first conditions in (\ref{currsimm}) give rise to the following equations
\eqn
& & 0=\lambda^k {f_{ki}}^j M^\cdot_{\ j} \label{a} \\ 
& & 0={f_{ki}}^j M^\cdot_{\ j} +\lambda^j f_{kij}  M^\cdot_{\ *} \label{b} \\
& & M^i_{\ l} M^j_{\ m} \lambda^k f_{ijk} ={f_{lm}}^k M^*_{\ k} +
\lambda^k f_{lmk} M^*_{\ *} \label{1} \\
& & M^i_{\ \cdot} M^j_{\ m} \lambda^k f_{ijk} =
\lambda^k {f_{km}}^h M^*_{\ h} \label{2} \\
& & M^i_{\ *} M^j_{\ m} \lambda^k f_{ijk} = 0 \label{3} \\
& & M^i_{\ *} M^j_{\ \cdot} \lambda^k f_{ijk} =0 \label{4} \\
& & {f_{ij}}^k M^i_{\ l} M^j_{\ m} -\lambda^j {f_{ij}}^k
(M^\cdot_{\ l} M^i_{\ m} -M^\cdot_{\ m} M^i_{\ l}) =M^k_{\ j} {f_{lm}}^j
+M^k_{\ *} \lambda^j f_{lmj} \label{lm} \\
& & {f_{ij}}^k M^i_{\ l} M^j_{\ \cdot} -\lambda^j {f_{ij}}^k
(M^\cdot_{\ l} M^i_{\ \cdot} -M^\cdot_{\ \cdot} M^i_{\ l}) =-M^k_{\ j}
{f_{hl}}^j \lambda^h \label{l.} \\
& & {f_{ij}}^k M^i_{\ l} M^j_{\ *} -\lambda^j {f_{ij}}^k
(M^\cdot_{\ l} M^i_{\ *} -M^\cdot_{\ *} M^i_{\ l}) =0
\label{l*} \\
& & {f_{ij}}^k M^i_{\ \cdot} M^j_{\ *} -\lambda^j {f_{ij}}^k
(M^\cdot_{\ \cdot} M^i_{\ *} -M^\cdot_{\ *} M^i_{\ \cdot}) =0 \ .
\label{.*} 
\feqn
Eq. (\ref{a}) can be solved by setting $M^\cdot_{\ j} =\sigma \lambda_j$. Inserting into
eq. (\ref{b}) we find  $M^{\cdot}_{\ *} =-\sigma$. Similarly, equations (\ref{3}) and (\ref{4})
are satisfied by $M^i_{\ *} = \epsilon \lambda_i$ and from (\ref{l*}, \ref{.*}) we obtain
$\epsilon = \sigma$.

If we define
\eqn
N^i_{\ j} \equiv M^i_{\ j} +\sigma \lambda^i \lambda_j \quad , \quad 
M^{\cdot}_{\ \cdot} \equiv \nu \quad , \quad   M^{*}_{\ *} \equiv \mu \quad , \quad
M^{*}_{\ \cdot} \equiv \gamma
\feqn
the rest of equations (\ref{1}, \ref{2}, \ref{lm}, \ref{l.}) can be written as
\eqn
& & {f_{lm}}^j ( M^*_{\ j} + \mu \lambda_j ) = \lambda_k {f^k}_{in} N^i_{\ l} N^n_{\ m} 
\label{e1}\\
& & \lambda^k {f_{km}}^j ( M^*_{\ j} + \mu \lambda_j ) = \lambda_k {f^k}_{il} N^l_{\ m} 
M^i_{\ \cdot}  \label{e2} \\
& & {f^k}_{ij} N^i_{\ l} N^j_{\ m} =N^k_{\ n} {f^n}_{lm} \label{struttura} \\
& & {f^k}_{ij} N^i_{\ l} M^j_{\ \cdot} = N^k_{\ m} {f^m}_{lh} \lambda^h + \nu
\lambda^h {f_{hm}}^k N^m_{\ l} \ .\label{e4} 
\feqn
From the first equation, using eq. (\ref{struttura}) we find $M^*_{\ j} = \lambda_k N^k_{\ j}
- \mu \lambda_j$. Inserting this result into the second equation we obtain
$M^i_{\ \cdot} = \lambda^k N^i_{\ k} + \eta \lambda^i$ and the last equation imposes
$\eta = - \nu$. 
Therefore the solution to (\ref{a}--\ref{.*}) reads
\eqn
M^I_{\ J} =\left(
\begin{array}{ccc}
N^i_{\ j} -\sigma \lambda^i \lambda_j \quad  & 
\lambda^k N^i_{\ k} - \nu \lambda^i \quad  & \sigma \lambda^i \\
\sigma \lambda_j & \nu & -\sigma \\
\lambda_k N^k_{\ j} -\mu \lambda_j & \gamma & \mu
\end{array}
\right)
\feqn
with $N^i_{\ j}$ satisfying eq. (\ref{struttura}).
On the matrix $M$ we have still to impose the second constraint of (\ref{currsimm}) and
(\ref{confinv1}). From (\ref{confinv1}), using the explicit expression (\ref{killing}) for the
Killing metric we obtain as the only nontrivial condition 
\eqn
N^l_{\ i} h_{lm} N^m_{\ j} =h_{ij}
\label{struttura2}
\feqn
which implies $N^l_{\ i}$ to be invertible and to define the isometry
group of the Killing metric $h_{ij}$.
We can realize $N^l_{\ i}$ as an element of the original 
semisimple Lie group in the adjoint representation
\eqn
N^i_{\ j} =\{ e^{\theta^k \sigma_k} \}^i_{\ j} \ , \qquad
\{ \sigma_k \}^i_{\ j} =f^i_{\ kj} \ .
\label{Nmatrix}
\feqn
It is then easy to see that (\ref{struttura}) is automatically satisfied.

Finally, the remaining equations 
in (\ref{currsimm}) give rise to the following conditions
\eqn
& & \sigma (\sigma \lambda^2 +\sigma b -2\mu) =0 \\
& & 1+\nu \sigma \lambda^2 -\sigma \lambda_i N^i_{\ j} \lambda^j +\sigma b\nu
+\sigma \gamma -\mu \nu =0 \\
& & \lambda^2 (1+\nu^2 )-2\nu \lambda_i N^i_{\ j} \lambda^j +b\nu^2 +2\gamma \nu
=b
\feqn
which can be solved in terms of one free parameter.


\section{Solution to the constraints in the contracted case}
\label{app:solution2}

We now study the solutions to the constraints (\ref{currsimm},\ref{confinv1}) 
in the case of the contracted algebra (\ref{calgebra}) where $f_{ij}$ is still
given in terms of a vector $\lambda^k$ (see eq. (\ref{polarized})). In particular, 
we have ${f_{ij}}^* = \lambda^k f_{ijk}$ and ${f_{\cdot i}}^j = \lambda^k {f_{ki}}^j$.
  
For $\lambda \neq 0$ the corresponding equations are
\eqn
&& M^\cdot_{\ j} {f_{\cdot i}}^j =0 \\
&& {f_{ij}}^* M^\cdot_{\ *} =0 \\
&& M^l_{\ i} M^k_{\ j} {f_{lk}}^* ={f_{ij}}^* M^*_{\ *} \\
&& M^l_{\ \cdot} M^k_{\ i} {f_{lk}}^* ={f_{\cdot i}}^j M^*_{\ j} \\
&& M^l_{\ *} M^k_{\ i} {f_{lk}}^* =0 \\
&& M^i_{\ *} M^j_{\ \cdot} {f_{ij}}^* =0 \\
&& M^\cdot_{\ i}{f_{\cdot l}}^k M^l_{\ j} -M^\cdot_{\ j} {f_{\cdot l}}^k M^l_{\ i} ={f_{ij}}^* M^k_{\ *} \\
&& M^\cdot_{\ \cdot} M^l_{\ j} {f_{\cdot l}}^k -M^l_{\ \cdot} M^{\cdot}_{\ j} {f_{\cdot l}}^k ={f_{\cdot j}}^l M^k_{\ l} \\
&& M^\cdot_{\ *} M^l_{\ i} {f_{\cdot l}}^k -M^l_{\ *} M^{\cdot}_{\ i} {f_{\cdot l}}^k =0 \\
&& M^\cdot_{\ *} M^j_{\ \cdot} {f_{\cdot j}}^k -M^\cdot_{\ \cdot} M^j_{\ *} {f_{\cdot j}}^k =0 
\feqn
together with the isometry conditions
\eqn
&& \Omega_{ij} = M^k_{\ i} M^l_{\ j} \Omega_{kl} +b M^\cdot_{\ i} M^\cdot_{\ j} +M^*_{\ i} M^\cdot_{\ j} +M^\cdot_{\ i} M^*_{\ j} \\
&& 0 = M^k_{\ \cdot} M^l_{\ j} \Omega_{kl} +b M^\cdot_{\ \cdot} M^\cdot_{\ j} +M^*_{\ \cdot} M^\cdot_{\ j} +M^\cdot_{\ \cdot} M^*_{\ j} \\
&& 0 = M^k_{\ *} M^l_{\ j} \Omega_{kl} +b M^\cdot_{\ *} M^\cdot_{\ j} +M^*_{\ *} M^\cdot_{\ j} +M^\cdot_{\ *} M^*_{\ j} \\
&& b = M^k_{\ \cdot} M^l_{\ \cdot} \Omega_{kl} +b M^\cdot_{\ \cdot} M^\cdot_{\ \cdot} +2M^*_{\ \cdot} M^\cdot_{\ \cdot} \\
&& 1 = M^k_{\ \cdot} M^l_{\ *} \Omega_{kl} +b M^\cdot_{\ \cdot} M^\cdot_{\ *} +M^*_{\ \cdot} M^\cdot_{\ *} +M^\cdot_{\ \cdot} M^*_{\ *} \\
&& 0 = M^k_{\ *} M^l_{\ *} \Omega_{kl} +b M^\cdot_{\ *} M^\cdot_{\ *} +2M^*_{\ *} M^\cdot_{\ *} \ .
\feqn
Solving these equations as before one finds for the matrix $M$
\eqn
{M_{\phantom{}_{(0)}}}^I_{\ J} =\left(
\begin{array}{ccc}
N^i_{\ j} & r^i \quad  & 0\\
0  & \nu \quad &  0 \\
s_j  & -\frac \nu2 r^2   \quad & \nu
\end{array}
\right)
\label{Mcontracted}
\feqn
where $N^i_{\ j}$ is still required to be an isometry of the metric and the following 
conditions must be satisfied
\eqn
&& \nu^2 =  1 \ , \\
&& \nu s_j =-r_l N^l_{\ j} \ ,  \\
&& {f_{ij}}^* =\nu N^l_{\ i} N^k_{\ j} {f_{lk}}^* \ .
\label{condcontracted}
\feqn
Multiplying the last equation by $N^{-1}$ and exploiting the explicit realization for
$N$, eq. (\ref{Nmatrix}), it is easy to see that the last condition can be satisfied only if 
$\lambda_j$, the vector in terms of which ${f_{ij}}^*$ is defined,  
is an eigenvector of $N$ with eigenvalue $\nu$ or $-\nu$. In fact, 
the last equation can be rewritten as
\eqn
\lambda_k =\nu {f_k}^{ij} {\tilde f^l}_{ji} {N^t}_l \lambda_t
\label{condcontracted3}
\feqn
where ${\tilde f^l}_{ij}$ are the structure constants in the basis
$\tilde \tau_i  \equiv {N^j}_i \tau_j$, being $\tau_j$ the generators of the original 
semisimple algebra (see eq. (\ref{A})). We define the matrix
${C_k}^l ={f_k}^{ij} {\tilde f^l}_{ji}$ and consider a generic matrix $R$ of the form 
(\ref{Nmatrix}). Since the matrix $R$ is an isometry which 
also preserves the structure constants (see (\ref{struttura})) it is quite 
easy to show that the following chain of identities holds
\eqn
{R^s}_h {C_s}^k= {R^s}_h {f_s}^{ij} {\tilde f_{ij}}^k ={f_h}^{ij} {R^l}_i
{R^m}_j {\tilde f_{lm}}^k ={C_h}^s {R^h}_k \ .
\feqn
Therefore the matrix $C$ commutes with all the elements of the group expressed
in the adjoint representation. Being the group semisimple, this representation
is irreducible so that $C$ must be proportional to the identity
\eqn
{C_k}^l =\alpha {\delta_k}^l \ .
\feqn
Using the fact that the coefficients $f$ and $\tilde f$ generate two different basis
of a D-dimensional subspace of the space of ${\rm D}\times {\rm D}$ antisymmetric
matrices with scalar product generated by $h_{ij}$, it is easy to show that 
\eqn
\tilde f^i_{\ jk} =\alpha f^i_{\ jk}
\feqn
with $\alpha =\pm 1$. Therefore eq. (\ref{condcontracted3}) reduces to 
${N^k}_i \lambda^i =\pm \nu \lambda^k$.

\end{appendix}

\end{document}